**Understanding the soil water dynamics during excess and deficit rainfall conditions over the Core monsoon zone of India**

Mangesh M. Goswami[1,3], Milind Mujumdar[1], Bhupendra Bahadur Singh[1], Madhusudan Ingale[1], Naresh Ganeshi[7,8,] Manish Ranalkar[2], Trenton E. Franz[4], Prashant Srivastav[5], Dev Niyogi[6], R. Krishnan[1] and S. N. Patil[3]

[1]Indian Institute of Tropical Meteorology, Pune-411008.

[2]India Meteorological Department, Pune

[3]Kaviyatri Bahinabai Chaudhari North Maharashtra University, Jalgaon-425001

[4]School of Natural Resources, University of Nebraska-Lincoln USA.

[5]Remote Sensing Laboratory, Institute of Environment and Sustainable Development, Banaras Hindu University, India

[6] The University of Texas at Austin, Department of Geological Sciences, Department of Civil, Architectural and Environmental Engineering, Austin, TX USA.

[7]Department of Atmosphere, Ocean and Earth System Modeling Research, Meteorological Research Institute, Tsukuba, Japan

[8]Atmosphere and Ocean Research Institute, University of Tokyo, Chiba, Japan

**Abstract:**

Observations of soil moisture (SM) during excess and deficit monsoon seasons between 2000 to 2021 present a unique opportunity to understand the soil water dynamics (SWD) over core monsoon zone (CMZ) of India. This study aims to analyse SWD by investigating the SM variability, SM memory (SMM), and the coupling between the surface and subsurface SM levels. Particularly intriguing are instances of concurrent monsoonal extremes, which give rise to complex SWD patterns. Usually, it is noted that a depleted convective activity and persistence of higher temperatures during the pre-monsoon season leads to lower SM, while monsoon rains and post-monsoon showers support the prevalence of higher SM conditions. The long persistent dry spells during deficit monsoon years enhances the Bowen ratio (BR) due to the high sensible heat fluxes. On the other hand, the availability of large latent heat flux during excess monsoon and post-monsoon seasons tends to decrease the BR. This enhancement or reduction in BR is due to evapotranspiration (ET), which influences the SWD by modulating the surface subsurface SM coupling. The surface and subsurface SM coupling analysis for CMZ exhibits significant distinction in the evolution of wet and dry extremes. SM variations and persistence time scale is used as an indicator of SMM, and analysed for both surface and subsurface SM observation levels. Evidently, subsurface SM exhibits remarkably prolonged memory timescales, approximately twice that of surface SM. Furthermore, we dissect SWD linked to wet and dry extremes by analysing annual soil water balance (SWB). Our findings reveal augmented (diminished) ET during deficit (excess) years, subjected to a higher (lower) number of break events. In essence, our study underscores the significance of surface-subsurface SM observations in unravelling the intricate tapestry of SWD.

**Introduction** -

Soil moisture (SM) is a critical variable of the hydro-meteorological system having a significant influence on land-atmosphere interactions (Mallet et al., 2020), and land surface processes such as evaporation, infiltration, percolation, runoff, drainage, and storage (Yin et al., 2019). Studies have highlighted the impact of surface SM on the amount of net radiation at the land surface (Eltahir 1998), and the partitioning of available surface energy between the latent and sensible heat fluxes (Entekhabi and Rodriguez-Iturbe 1994; Rasheed et al., 2022). The land surface processes can be better understood by analysing the soil water dynamics (SWD) at the surface and subsurface levels (O'Geen 2013). Rainfall is the primary factor that affects SM variability, particularly over the regions dominated by monsoons such as the Indian subcontinent (Ganeshi et al., 2020). The longer duration monsoonal rain spells with concurrent cooler temperature tends to reduce the atmospheric water demand (reduced evaporative loss) leading to an increase in surface and subsurface SM content (Kemp, 1983). Contrary to this, non-monsoonal short-duration rain spells with concurrent warmer temperatures tend to enhance the boundary layer atmospheric water demand (enhanced evaporative loss), which promotes a net loss of SM content in surface and deeper layers (Kumar et al., 2009). These variations in the atmospheric water demand during the monsoon and the post-monsoon months can enhance or suppress convective / thunderstorm activities during the subsequent pre-monsoon season (Abera et al., 2020).

Koster et al., (2004) emphasized the role of subsurface SM variations on the land-atmosphere coupling strength by bringing out the crucial role of SWD in modulating the coupling. However, the present understanding about the connection between surface and subsurface SM variations and their relationship with soil temperature (ST) to get insights into the SWD remains relatively low. (Xu et al., 2022). It is also known that the root zone SM may affect the movement of water and energy between the surface and the atmosphere, and it has a memory that could be harnessed to improve weather forecasting (Kumar et al., 2009). The evolution of hydrological extremes can be assessed from SM profiles and the corresponding SWD (Famiglietti et al., 2021). Surface–subsurface SM coupling, along with the SM memory, is an important element of the region-specific physical processes, an understanding of which may lead to a better understanding of the consequences of regional climate change (Singh et al., 2021).

Various factors like soil characteristics (Baroni et al., 2013), vegetation, geographical location (Qiu et al., 2001), solar radiation (Western and Blöschl, 1999),  climate (Montenegro and Ragab, 2012), and land use factors (Venkatesh et al., 2011) affect the SM variability (Zarlenga et al., 2018). Also, the degree of SM variability depends on the hydrological fluxes and the initial SM state (Albertson & Montaldo, 2003). Over central India, the spatial distribution of  SM and rainfall variations are usually consistent despite some discrepancies arising due to other regional factors (Sathyanadh et al., 2016). The SM variability at different depths is crucial to understand the association between SM and ST (Hirschi et al., 2014). The importance of surface SM variations in seasonal wet and dry conditions during the monsoon period was illustrated by Varikoden and Revadekar (2018). SM variations are also known to influence the persistence and amplification of hydro-meteorological extremes through land–atmosphere interactions (Ferranti & Viterbo 2006; Ganeshi et al., 2020). However, most of the studies that focus on the land-atmosphere coupling fall short of addressing the role of subsurface SM.

Earlier studies showed that the spatiotemporal distribution of SM significantly changes with soil depth (Zhang & Huang, 2021). Unlike rainfall variability, surface SM variability prevails throughout the year (Albergel et al., 2013), whereas the subsurface SM majorly varies on seasonal scales despite minor fluctuations in-between (Peterson et al., 2019). Daily variations in surface and subsurface SM are crucial for estimating the inflow (infiltration) and outflow (evapotranspiration (ET) and deep drainage) of water fluxes and water mass balance (Franz et al., 2012). In this context, exploring surface-subsurface SM coupling can constrain the uncertainties in estimating water and surface energy balance. Xu et al., (2022) emphasised the importance of surface and subsurface SM coupling to estimate the subsurface SM. In light of the above discussions, it is important to understand level-by-level SM variability. Apart from the SM variability as discussed above, the soil also has a unique characteristic of long-time remembrance of wet or dry conditions (known as soil moisture memory – SMM) induced by the atmosphere and land surface processes (Delworth and Manabe, 1988; Seneviratne et al., 2006a). The long-term variation in tropical rainfall modulates the SMM time scale through terrestrial hydrological processes (Asharaf & Ahrens, 2013).

Land-atmosphere coupling can be realistically evaluated by analysing SMM, along with other elements of the climate system, for their better representation in climate models (Seneviratne et al., 2006b, Koster et al., 2010). Several observational and modelling studies have demonstrated the latitudinal dependence of SMM and found it to be relatively higher in the extra tropics than in lower latitudes (Manabe and Delworth, 1990; Wang et al., 2010). Persistent rain spells during the monsoon season can induce an extended time scale of SMM. The highest surface SMM is found over wet regions of the Western Ghats, north-east, and northern parts of India (Ganeshi et al., 2020). It is to be noted that earlier studies have dealt with surface SMM (e.g., Albertson and Montaldo, 2003); however, the importance of subsurface SMM remains unexplored. Interestingly, across the tropics, the evaluation of surface SMM is more complex, particularly over the core monsoon zone (CMZ) (Dong & Ochsner, 2018; Ganeshi et al., 2023). Inevitably, subsurface SMM may also exhibit complex behaviour. Due to the sparse availability of subsurface SM observations, most of the earlier research was confined to the surface SM based inferences which hindered the understanding of SWD. Thus surface and subsurface SM observations are crucial to get better insights into the SWD. In this study, we have examined SM and hydro-meteorological datasets from multiple sources such as IMD, Cosmic ray soil moisture observations system at Indian Institute of Tropical Meteorology (COSMOS-IITM) over the field scale, in-situ observations, reanalysis, and model data products. The paper is organized as follows: section 2 discusses data and methodology, section 3 describes hydro-meteorological features on inter-annual, seasonal, and sub-seasonal scales over the study region, followed by the results based on analysis of surface and subsurface coupling, SWD, SMM, and soil water balance (SWB) using various data sets. The conclusion is presented in section 4.

## 2) Data and Methodology:

### 2.1) Data:

The study area consists of the Indian core monsoon zone (CMZ), which encompasses latitudes between 17.0° to 30.1°N and longitudes of 71.5° to 88.7°E (see Figure 1, Rajeevan et al 2010). This study incorporates in-situ SM measurements taken by the India Meteorological Department

(IMD) at selected agro-meteorological weather stations across the CMZ of India. These measurements are available at weekly temporal resolution and at different depths (10, 20, 30, 40, 50, and 60 cm). Owing to the significant data gaps at several locations, we selected 8 stations that qualify for at least 60% quality check. (Stations with large data gaps (> 4 weeks) and / after dropping SM values exceeding 10% of the field capacity were removed. Whereas, small gaps were replaced by the climatological values).

We also took into account daily gridded precipitation (Pai et al., 2014) and surface air temperature (https://www.imdpune.gov.in/cmpg/Griddata/Max_1_Bin.html) data from IMD for the period of 2000 to 2021, along with various hydro-meteorological observational datasets at the field scale from the COSMOS-IITM site (Ganeshi et al., 2020; Mujumdar et al., 2021). Furthermore, we incorporated various hydro-meteorological observational datasets at the field scale from the COSMOS-IITM site (Ganeshi et al., 2020; Mujumdar et al., 2021) for the years 2017 to 2022. These datasets are based on COSMOS, a combination of standard sensors (such as Acclema TDR probe, Stevens hydra probe, Theta Probe), low-cost sensors (capacitive sensors assembled and calibrated at the IITM-COSMOS site), and the gravimetric method. In addition, the reanalysis data products from the European Centre for Medium-Range Weather Forecasts (ECMWF) (ERA-5), the National Centre for Medium-Range Weather Forecasting (NCMRWF), the Indian Monsoon Data Assimilation and Analysis (IMDAA), Land Data assimilation system (LDAS, Chen et al 2007) and the Global Land Data Assimilation System (GLDAS; Rodell et al., 2004)) are used in this study (for more details see Table 1). These in-situ observations along with the reanalysis, and model data products were utilised to examine the study goal for a common period of 2000-2021. Given the limited availability of soil moisture (SM) observations, we performed statistical assessments on various data products, including ERA-5, IMDAA, LDAS, and GLDAS, spanning the years 2000 to 2021.

**2.2) Methodology:**

In this study, we evaluated and analysed various data products to estimate the intra-seasonal and inter-annual variability of SM and rainfall. SM values at 10 cm depth are considered as the surface SM and that of at 60 cm refers to subsurface SM. Surface-to-subsurface SM coupling analysis is then carried out to infer the impact of surface SM variance on the subsurface, as well as on the SMM time scales at various depths using the autocorrelation and soil water balance analyses. A monthly climatology and the corresponding monthly anomaly is calculated for the period of 2000 to 2021.

**2.2.1) Analytical techniques:**

Understanding the layer-by-layer SWD is crucial to determine the surface-to-subsurface SM connection (Xu et al., 2022; Malik et al., 2021; Hirschi et al., 2014). This can best be done by the coupling strength analysis. Earlier studies explored the coupling strength between surface SM and surface air temperature using several analytical techniques and model simulation experiments (Dong & Crow, 2018; Miralles et al., 2012; Seneviratne et al., 2013; Ganeshi et al., 2023). Similarly, the coupling strength between surface (10 cm) SM and subsurface (60 cm) SM was explored using the cross-correlation analysis (Xu et al., 2022; Tian et al., 2020). In the present study, we implemented following two approaches – (1) Regression analysis (Dirmeyer, 2011) and (2) Cross-correlation analysis (Xu et al., 2022). In the first approach, for each grid point, linear regression of weekly surface SM on subsurface SM anomalies is evaluated to estimate the slope $R_c$ for each year during 2000 - 2021. To account for the influence of SM variability on the slope (indicative of the strength of correlation), we also computed the standard deviation of weekly surface SM ($\sigma_{SM}$) for each year across the study period (Table 2). The coupling strength is therefore given by (Dirmeyer, 2011)

$$\Omega = \sigma_{SM} \times R_c \qquad (1)$$

Higher positive values of $\Omega$ indicate the regions where surface SM has a dominant impact on subsurface SM variability. On the other hand, lower values point toward the region where this impact is insignificant. Second approach focuses on cross correlation analysis and characterizes lead/lag relationship between the surface and subsurface SM at different depths: (Georgakakos et al., 1995; Mahmood & Hubbard, 2007). We computed the cross-correlation between weekly anomalies of surface and subsurface SM. The analysis is carried out for different lag values for each year during the study period. The maximum cross-correlation coefficient is indicative of the strength of the coupling and the sign of the lag corresponding to the maximum correlation highlights the potential of predicting subsurface from the surface SM.

Root mean square error (RMSE) and Unbiased root mean square error (UbRMSE) are mainly used to define the level of agreement between the surface and subsurface SM for the different data sets. The bias can be easily eliminated by identifying the UbRMSE, which characterises random error in order to obtain a reliable estimate of RMSE (Albergel et al., 2013; Brocca et al., 2013; Wu, 2016; Sathyanadh et al., 2016; Watterson, 2008; Wilks, 2011; Mujumdar et al., 2021). Furthermore, we implemented probability density function, Taylor statistics (Taylor 2005, 2001), and associated statistical analysis to understand the variation in soil temperature (ST) and SM at surface and subsurface levels.

## 2.2.2: Classification of Deficit and excess years:

The seasonal climatology of the monsoonal rainfall ($\bar{R}$) over CMZ of India is computed for the period of 2000 to 2021. The corresponding seasonal anomaly ($r_a$) for each year is then expressed as a percent deviation of mean (climatology) using equation (2),

$$\text{Percent deviation of mean} = \frac{r_a}{\bar{R}} \times 100 \text{ \%} \qquad (1\text{2})$$

The year with a percent deviation of mean > 10 % is called as an excess year. When it is less than -10 % we define it as a deficit year, otherwise, it is a normal year (see also supplementary Figure S1). Our definition is consistent with that of the drought and flood years based on the all-India area weighted summer monsoon rainfall anomaly (see also: https://mol.tropmet.res.in/monsoon-

, Mujumdar et al., 2023, Krishnan et al., 2020). Singh et al., (2018) have implemented similar criteria over central India to identify the drought and flood years.

## 3) Results and Discussion:

### 3.1) Analysis of hydro-meteorological parameters over the CMZ of India

#### a) Rainfall variability:

Understanding the spatiotemporal variability of various hydro-meteorological variables such as rainfall, SM, and temperature over the CMZ of India is crucial to get insights into the surface and subsurface hydrological processes over the region. Climatological mean summer monsoon rainfall computed over the period of 20 years over the CMZ is about 964 mm with a standard deviation of 132 mm. Figure 2(a) shows the area-averaged time series of monthly rainfall anomalies over the CMZ of India for the period 2000-2021. Considering the monsoon months (June – September), we have selected six deficits (2000, 2002, 2004, 2009, 2014, 2015) and six excess (2003, 2006, 2011, 2013, 2019, 2020) years based on the analysis carried out over the CMZ using the IMD gridded data (see Figure S1 and section 2.2.2; Singh et al., 2018).

Our analysis indicated that occurrences of wet spells (rainfall > 2 mm for 3 or more consecutive days) followed by break spells were relatively more (less) during the deficit (excess) years. Similar rainfall analysis is also carried out at the observational site located at the IITM, Pune campus and is also found to be consistent with the CMZ. The impact of deficit and excess rainfall during these years has a bearing on the surface and subsurface SM variability and affects the SWD of the region. This in turn influences the surface and subsurface SWD, SWB, and SMM time scales (discussed in detail in sections 3.3 onwards). The surface air temperature also plays a vital role in the modulation of hydrological extremes; therefore, its relationship with SM needs to be explored appropriately.

#### b) Surface air temperature variability:

The surface air temperature exhibits latitudinal variations, where the temperature is lower at higher latitudes and tends to increase gradually as we move towards the equator (Ganeshi et al., 2020). The monthly mean surface air temperature across the CMZ region exhibits strong seasonal variability, reaching its maximum (> 33 °C) during the pre-monsoon months and minimum (< 18 °C) during the post-monsoon months (Zhou et al., 2019). Area averaged monthly temperature anomaly time series for the period of 2000-2021 is shown in Figure 2 (b). Among the six deficit years, the monthly mean temperature variability (standard deviation) in pre-monsoon season is higher for 2002 (3.9 °C) while lower in 2009 (2.0 °C)

(see:https://mausamjournal.imd.gov.in/index.php/MAUSAM/issue/archive). Similarly, during the monsoon season, the surface air temperature variability was found to be higher in 2019 (2.4°C) and lower in 2002 (1.7°C). Additionally, we carried out a similar analysis over the COSMOS-IITM, Pune, India. It is worth mentioning that the extreme temperature events (daily maximum 2m temperature > 40°C and persisting for three or more consecutive days increased significantly during 2017-2022 and are consistent with results over CMZ (see Table S1). Additionally, in order to comprehend the impact of temperature change during the deficit and excess years, it is also required to analyse the variation in SM.

**c) Surface and subsurface SM variability:**

The monthly mean surface (10 cm) and subsurface (60 cm) SM variations are represented using the Box and whisker plot for the period of 2000-2021 over the CMZ of India in Figure 2(c) and 2(d), respectively. Usually, surface SM exhibits higher variability. Our statistical analysis shows that the coefficient of variation for SM at the surface level is about 49% in the pre-monsoon season, which increases to 61% during the monsoon season. However, the subsurface SM variability is found to be higher in extremely dry seasonal conditions (e.g., 2019 pre-monsoon; 65%) than that on concurrent surface SM (49%). Interestingly, during the 2019 pre-monsoon (void of convective activities) lower SM variability (9.8%) is found at the surface than that of the subsurface (13.4%) over the IITM-COSMOS site (see Table S2). Furthermore, we validated the surface and subsurface SM data products with the in-situ IMD observations to understand the variability of reanalysis and model data products.

**3.2) Validation of various data sets with the IMD observations across the CMZ.**

A cross-validation of the observations and reanalysis products is summarised in this section. It is intriguing to see how various SM data products can depict the monsoon's onset and withdrawal phases. IMD in-situ SM data products are used in this study as the "ground truth" to validate a range of available coarser-resolution data products. This subsection examines the surface and subsurface SM products from LDAS, GLDAS, ERA5, and IMDAA with IMD observations across the CMZ over the period 2000 to 2021 on a weekly time scale (Figure 3). Furthermore, we used the ERA-5, IMDAA, LDAS, and GLDAS data set to conduct a comparative study at the COSMOS-IITM site.

The variances observed at the Pune location are in line with the CMZ, and the ERA-5, IMDAA, and GLDAS data sets exhibit reasonable surface SM variability, suggesting realistic summer monsoon impacts from 2017 to 2022 (Figure S2). It is clear that the Indian summer monsoon, which spans the period of around the end of May to the start of October, plays a significant role in the seasonal changes in SM.

Further, we utilize Taylor statistics (Taylor et al., 2005, 2001) for calibration and validation of surface and subsurface SM on weekly observations against coarser-resolution reanalysis data products (ERA-5, IMDAA, LDAS, and GLDAS) over the CMZ of India (see Figure 3(a) and 3 (b)) and the Pune site (Figure S3). All reanalysis datasets considered in this study seem to correlate well with observations, however, the GLDAS product is found to have the best compatibility with the observations. The RMSE value in the case of GLDAS is noted to be 3.9 % and 4.2 % at the surface and subsurface level, respectively. In contrast, the IMDAA and ERA-5 and LDAS show an RMSE of 5.7 % (5.1 %) ,5.0 % (5.1 %) and 7.7% (6.0%) for the surface (subsurface) level, respectively. Overall results indicate how closely GLDAS SM data product correlates with the observational data and it is verified in the recent work by Ganeshi et al., (2020) and Mujumdar et al., (2021) over the COSMOS-IITM site. Furthermore, we are analysing the surface-subsurface SM coupling to get insight into the SWD.

### 3.3) Surface and subsurface SM coupling:

The variability of the surface SM is strongly influenced by atmospheric conditions throughout the year (Zheng et al., 2022), while the subsurface SM is more influenced by seasonal changes (Hirschi et al., 2014). Figure 4(a) and Figure 4(b) shows the surface (10cm) and Figure 4(c) and Figure 4(d) shows subsurface (60 cm) SM variability during the deficit (2002) and excess (2019) years respectively. The surface–subsurface linkage is analysed for the period 2000-2021 using the two approaches such as cross-correlation analysis and the method described by Dirmeyer (2011). This study estimates the coupling strength (see Table 2) at a maximum latency of 52 weeks (or one year). Shumway and Stoffer (2010) highlighted that the highest cross-correlation with positive latency indicate surface SM precedes subsurface SM. Whereas, negative latency shows the dominance of subsurface SM. We carried out a similar analysis and found that during an excess (2019) year a maximum cross-correlation (0.7) is associated with a negative lag of ~ a week (Figure 4f). Whereas the deficit (2002) year yielded a maximum cross-correlation at 0.5 with a positive lag of ~ 2 weeks (Figure 4e). These results during the deficit and excess years describe the importance of subsurface SM in the hydrological processes.

On an average, both the surface and subsurface become saturated (unsaturated) during an excess (deficit) year, for example, the 2019 (2002) monsoon. At this point, infiltration decreases (increases) and runoff increases (decreases). In other words, during the excess (deficit) year, the subsurface (surface) SM drives the surface (subsurface) SM. Also, as the persistence time scale of the subsurface is more (less), for the excess (deficit) case, the saturated (unsaturated) subsurface can (cannot) drive the variability at the surface, and as an aggregate effect it appears to be leading

(lagging) with a negative (positive) lag in Figure 4 (e, f). Additionally, surface-subsurface coupling strength was estimated using the ERA-5, IMDAA, LDAS, and GLDAS data sets for inter-comparison (Figure S4). The results show a significant bias (ERA5=17%, IMDAA= 19%, LDAS = 23%, and GLDAS=21%) in the coupling between surface and subsurface SM as compared to the observations, which underlines the importance of the observations. Furthermore, it could be intriguing to analyse the SM memory to get further insights into SWD in the context of surface and subsurface SM coupling.

**3.4) Analysis of SM memory:**

In the present study, we have analysed surface and subsurface SMM over CMZ using the autocorrelation function. The autocorrelation of weekly SM data products is calculated with a lag of 52 weeks (annual) for the period of 2000-2021 (Figure 5). In addition, we have estimated the SMM for the previously identified deficit and excess years shown in Figures 5(a) and 5(b), respectively. Here, the dotted green line (at 0.2) indicates the 95% significance level for the surface and subsurface SMM. The results show that, in general, the memory at the surface is confined to ∼6-7 weeks, while at the subsurface, it is extended to ∼8-9 weeks. However, during deficit years, the surface and subsurface SMM decreased to ∼2-3 and ∼4-5 weeks, respectively. Moreover, in the excess years, the SMM increased at both the surface and subsurface levels for 8-9 and 10-11 weeks, respectively. Additionally, the overestimation of SMM obtained using the reanalysis (ERA-5, IMDAA) and model data (GLDAS, LDAS) products over the CMZ during the deficit as well as excess years (see Figure S5), highlights the need of soil moisture observations.

In combination with Figure 2, Figures 5, 4(e), and 4(f) show a changing pattern in subsurface SM as a response to surface SM during deficit years, which corresponds to reduced coupling strength (lower value of maximum correlation, Figure 4) between surface and subsurface SM. This in turn reflects in the reduction of SMM. Whereas during excess years, the subsurface SM is found to be more influenced by the surface SM which enhances the coupling strength (higher correlation value – Figure 4f) between surface and subsurface SM. This is aptly reflected in the SMM during excess years. The variation in the SMM may affect the surface energy exchanges, which are conducive to the initial development of the thunderstorm/convective system over the land. Therefore, it would be worth exploring the role of surface fluxes with the concurrent SM variations to get insight into SWD.

**3.5) Land-atmosphere energy exchanges in successive seasons:**

The surface soil temperature (ST) has a significant impact on ET, and it affects the earth's water cycle as well as the energy balance between the land and atmosphere (Hinkel et al., 2001). The evapotranspiration rate is largely controlled by SM availability at the surface until it fulfils the atmospheric water demand. Eltahir (1998) suggested the SM-rainfall feedback mechanism based on the control of SM on surface albedo and Bowen ratio (BR). He emphasized that both factors decrease with an increase in the surface water content, which has a direct effect on the land-atmosphere energy exchanges. In general surface SM interacts with temperature through its effects on the sensible heat flux (SHF) and latent heat flux (LHF). Interestingly, in the dry seasonal conditions due to sensible heat flux, land-atmosphere energy exchanges are relatively higher (Ge et al., 2019). Therefore, we analysed the pre-monsoon land-atmosphere energy exchanges by comparing the BR (the ratio of SHF with LHF, obtained using GLDAS data set) with the associated rainfall anomalies for the period of 2000-2021 over the CMZ (Figure 6).

In the previous section 3.4, we have shown that the SMM at the surface and subsurface level is higher (lower) in the excess (deficit) years, which implies the sufficient (scarce) availability of moisture for the land-atmosphere interaction. The wet conditions during excess year, which persists for the post monsoon months as positive rainfall anomalies (Figure 6), can extend to the winter and pre-monsoon seasons of the succeeding year. Wet SM conditions, as is pointed out by the Eltahir (1998), reduces the BR by regulating the turbulent heat fluxes. This favours the increase in atmospheric water vapor content which in turn enhances the convective activities in the winter and pre monsoon months of the successive year (Figure 6). Contrary to this, deficit monsoon year is observed to be dry (negative rainfall anomalies) during post monsoon months. SMM at the surface and subsurface level hence reduces rapidly, leading to the scarcity of surface SM. This may lead to fall short of the atmospheric water demand by increasing the BR and thus, sensible heat flux. Resulting void of convective activities is noticed during pre-monsoon of successive year of the deficit years in Figure 6. Similarly, we have also analysed the impact of SM and ST variability on the BR over the IITM-COSMOS site for the period of 2019-2022 (see Figure S6). We noted that the enhanced BR during the pre-monsoon season is conducive for decreased surface SM and ST variability. Therefore, it would be worth exploring the role of SM variability with the concurrent ST variations at subsurface level to get insight into the SWD.

**3.6) Insights into the soil water dynamics:**

Due to the spars availability of surface and subsurface SM and ST data over the CMZ region, we have computed the monthly surface and subsurface ST and SM distribution using daily data sets at the COSMOS-IITM site for the period of 2019-2020 (see Figures (7a) and (7b)). We have arranged the data set (2019-2022) in ascending order and computed values for the 5th and 95th percentile of this data. The dotted red lines indicate the threshold for extremely dry (warm) SM (ST) conditions (corresponding to 95th percentile of the data), whereas the dotted blue line indicates the threshold for extremely wet (cool) SM (ST) conditions (5th percentile of the data).

Our results show that absence of thunderstorm/convective activities during the pre-monsoon season of 2019 and 2022 enhances extreme temperature conditions over Pune region, clearly depicted by the remarkably deficit surface and subsurface SM and higher ST exceeding the 95th percentile threshold. On the other hand, surface and subsurface SM (ST) are confined to a much higher (lower) threshold due to enhanced convective activities during the 2020 and 2021 pre-monsoon seasons. Interestingly, the surface ST is much lower during the 2020 monsoon season, while subsurface ST is confined well below 25 °C during the monsoons of 2020 and 2021 (Figure 7a).

**3.7) Analysis of soil water balance:**

Analysis of soil water balance serves as an important tool in understanding hydrological processes. Earlier studies have explored the soil water balance over the floodplain region, where the soil water dynamics is controlled by the stream flow, precipitation, and groundwater (Gabiri et al., 2018; Jain, 2012). Numerous studies have highlighted the importance of appropriately accounting for precipitation and ET when evaluating the soil water balance across different regions of India. In this study, following Franz et al., (2012), we have estimated the net soil water balance for surface (L1= 10 cm) and subsurface (L4= 60 cm) levels using field-scale in-situ observations, over the COSMOS-IITM site (Figure 8) for the period of 2019-2022. The Blue bar indicates the net infiltration (Inflow) while the red bar shows net ET and deep drainage (Outflow). The difference between inflow and outflow shows the net storage (grey line) for the surface and subsurface layers. Interestingly, the higher storage at the surface during 2020 can be attributed to intense monsoon rain spells during the excess monsoon and post-monsoon rainfall of 2019. However, the strong pre-monsoon thunderstorm/convective activities, monsoon break events, and moderate rain spells during 2021 led to enhanced net ET (4.69 mm/day), causing reduction in surface storage. It is worth mentioning here that the net estimation of the above-estimated fluxes is consistent with those of in-situ Eddy covariance observations. The above estimation using reanalysis and model data products depicts an unrealistic estimation of the net storage, infiltration, and fluxes (see Table S3).

**4) Conclusion:**

In this paper, we have studied the surface and subsurface soil moisture (SM) variability over the core monsoon zone (CMZ) of India. This study uses surface and subsurface IMD observations along with various hydro-meteorological data products such as GLDAS, ERA5, LDAS, and

IMDAA for the period 2000-2021. Additionally, various field-scale hydro-meteorological observations from the COSMOS-IITM site during the period of 2017 – 2022 are implemented.

The result indicates that the SM variability at the surface is higher throughout the year, while the subsurface variability is higher during extremely dry or wet conditions. The validation statistics, using the Taylor diagram, brings out the reasonable agreement of ERA5, IMDAA, LDAS, and GLDAS SM products with that of the IMD observations. The coupling strength between the surface and subsurface is estimated for the study period of 2000-2021. The coupling strength was higher (~ 8.7) for the excess year (2019) than that of for the deficit year (~0.9, in 2002). The increase or decrease in the coupling strength depicts the significance of SMM under extremely dry and wet conditions. The results overall indicate that the SMM is about 6-7 weeks at the surface, while it is 9-10 weeks at the subsurface level for the period of 2000-2021. Interestingly, in deficit years, the SMM at subsurface decreases by ~50%, however, in the excess years, it increases by ~7-10 %. The enhanced and suppressed SMM during the Indian summer monsoon season has a significant impact on the land-atmosphere energy exchange.

Our findings indicate that the enhancement (reduction) in rainfall during the Indian summer monsoon may lead to an increase (decrease) in the SMM at the surface and subsurface level during the succeeding post-monsoon season. The increase in SMM implies the sufficient availability of moisture for the land-atmosphere interaction in the successive pre-monsoon seasons. However, during deficit monsoon years less SM is available at both surface and subsurface levels, consequently energy exchange is weaker (reduced latent and increased sensible heat flux). This reduction in the land-atmosphere energy exchange hinders the development of convective / thunderstorm activities. On the contrary increased energy exchange, during excess monsoon conditions (higher latent heat flux) has a direct effect on modulating the soil water dynamics by decreasing the SM and increasing the ST variability. Which in turn supports thunderstorm / convective activities in succeeding pre-monsoon season. Furthermore, the surface and subsurface

SM variability underlines the importance of soil water balance during wet and dry extremes. Our results highlighted that the storage (recharge) at the surface is comparatively positive during 2020, which could be attributed to intense monsoon rain spells following the excess monsoon and post-monsoon rainfall during 2019. However, in the deficit year recharge period is higher with an enhanced net ET (e.g., 4.69 mm/day in 2021), due to longer duration of monsoon break events and moderate rain spells. Our work emphasizes the need to investigate the role of ET in the variation of surface and subsurface SMM for dry and wet extreme cases in the future. Additionally, in view of agricultural applications, it is worth exploring the impact of thermal gradient on SWD.

**Tables:**

**Table 1: Observation, reanalysis, and model data products used for analysis and validation.**

| Type of data | Name of dataset / Source | variables | Spatial Resolution | Temporal Resolution | Period | References |
|---|---|---|---|---|---|---|
| **A ) Gridded Data Sets** | | | | | | |
| Observation | IMD | Rainfall | 0.25 x0.25 | daily | 2000-2021 | Pai et al., 2014 |
| | IMD | Temperature | 1 x 1 | daily | 2000-2021 | Basha et al., 2017, Gupta et al., 2020 |
| Reanalysis | IMDAA | Soil moisture, Soil temperature, Latent heat flux, sensible heat flux, evapotranspiration | 0.12 x0.12 | daily | 2000-2021 | Rani et al., 2021 |
| | ERA-5 | | 0.25 x 0.25 | daily | 2000-2021 | Chen et al., 2007 |
| Model Data | GLDAS | | 0.25 x0.25 | daily | 2000-2021 | Syed et al., 2008 |
| | LDAS | Soil moisture, Soil temperature | 0.04 x 0.04 | daily | 2000-2017 | Nayak et al., 2018 |
| **B) In situ Observational data sets** | | | | | | |
| observation | IMD (8 stations Over CMZ of India) | Soil moisture | In-situ | Week | 2000-2021 | (Sathynadha et al., 2016) |
| | IITM-COSMOS | Soil moisture, Soil temperature, pressure, relative humidity, rainfall, surface temperature | In-situ | daily | 2019-2022 | Majumdar et al., 2021 |

**Table 2: Rainfall Anomaly and coupling strength (Dirmeyer 2011) for the period of 2000-2021. The deficit and excess years are indicated by Red and Blue colour respectively.**

| Year | CMZ Anomaly | Coupling |
|------|-------------|----------|
| 2000 | -18.6 | 4.2 |
| 2001 | 1.2 | 2.0 |
| 2002 | -23.3 | 0.9 |
| 2003 | 11.9 | 1.4 |
| 2004 | -14.2 | 1.0 |
| 2005 | -0.6 | 2.6 |
| 2006 | 17.8 | 3.0 |
| 2007 | 7.0 | 4.5 |
| 2008 | 2.5 | 2.4 |
| 2009 | -19.9 | 0.7 |
| 2010 | 0.9 | 2.9 |
| 2011 | 16.0 | 3.3 |
| 2012 | 0.4 | 2.0 |
| 2013 | 21.2 | 3.4 |
| 2014 | -11.8 | 1.5 |
| 2015 | -11.0 | 0.7 |
| 2016 | 6.0 | 1.6 |
| 2017 | -8.9 | 3.6 |
| 2018 | -9.2 | 3.0 |
| 2019 | 19.8 | 8.7 |
| 2020 | 10.0 | 7.7 |
| 2021 | 4.4 | 6.7 |

**Figures:**

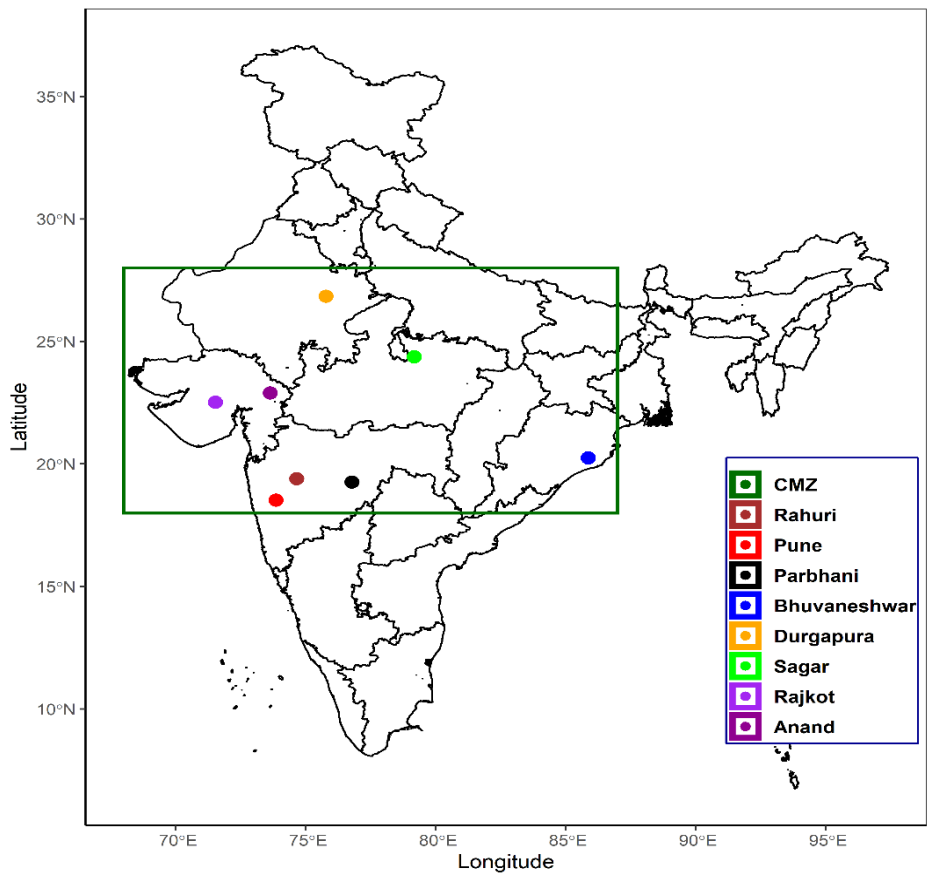

**Figure.1:** Various IMD Hydro-meteorological observation stations across the core monsoon zone (CMZ) of India (represented by distinct colour inside the CMZ box).

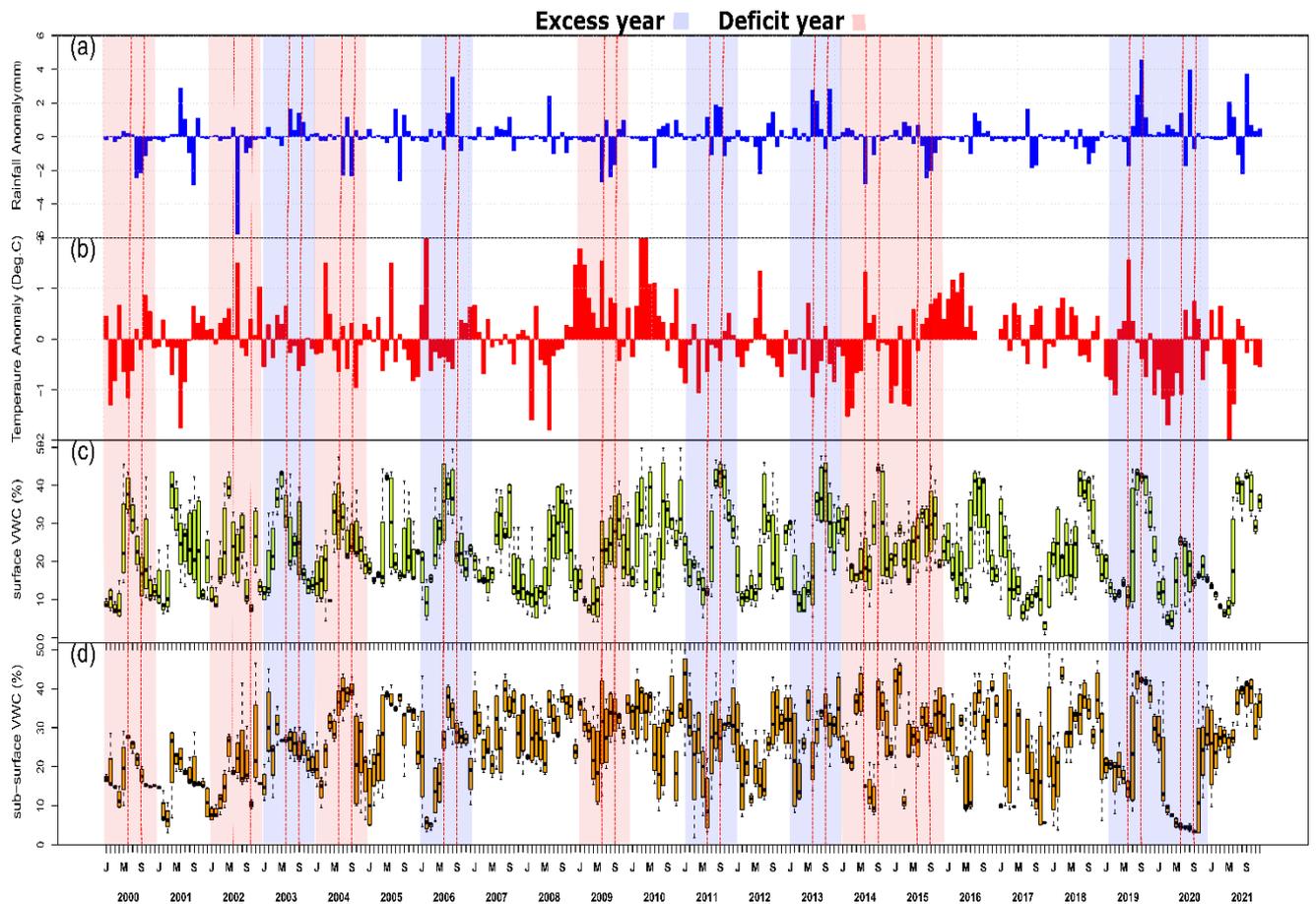

**Figure. 2:** Time series of monthly (a) rainfall anomaly (mm) and (b) 2M temperature anomaly (°C). Monthly statistics represented as Box and Whisker plots for weekly (c) surface, and (d) subsurface volumetric SM (%) content based on IMD in-situ observations over the CMZ of India for the period 2000–2021. The dotted red line indicates the monsoon months (JJAS). The shaded area in red and blue indicates the deficit and excess years respectively.

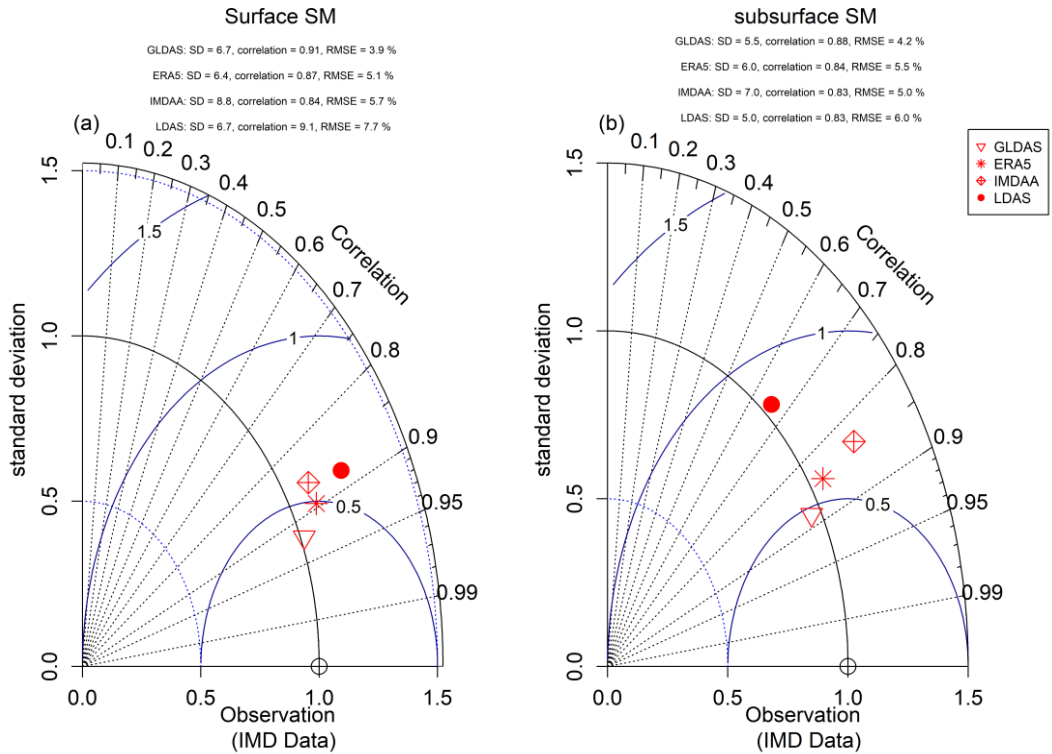

**Figure 3:** Taylor Diagram shows a) surface (10cm) SM and b) Subsurface (60cm) SM comparison of four different SM data sets (ERA-5, IMDAA, LDAS and GLDAS; represented by distinct symbols) with validation reference to IMD observations in terms of centered Root Mean Square Error (RMSE), Correlation Coefficient, and Standard Deviation.

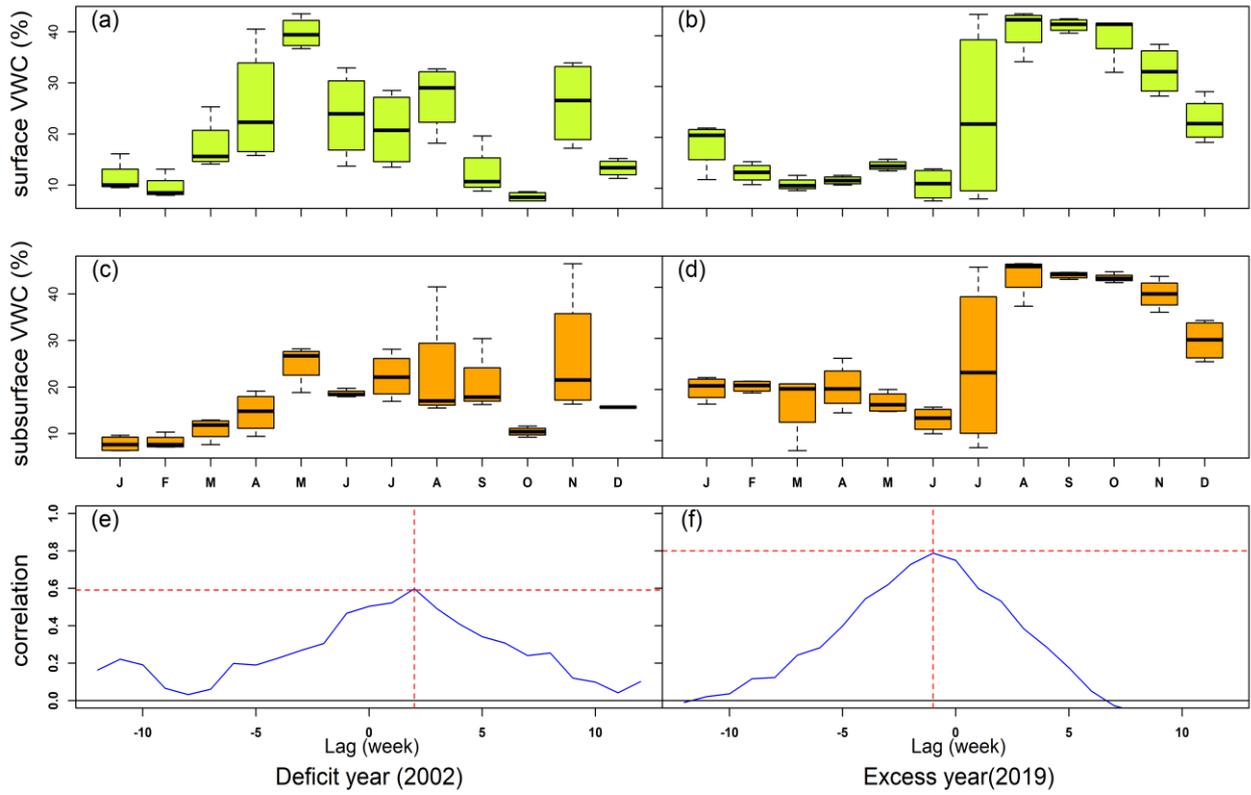

**Figure 4:** Upper and middle panels represent monthly surface and subsurface SM variability respectively as a Box and Whisker plot. Last panel shows lag-correlation between surface and subsurface SM. All the left panels (a), (c) and (f) are for the deficit year (2002). Whereas all the right panels (b), (d) and (f) are for the excess year (2019).

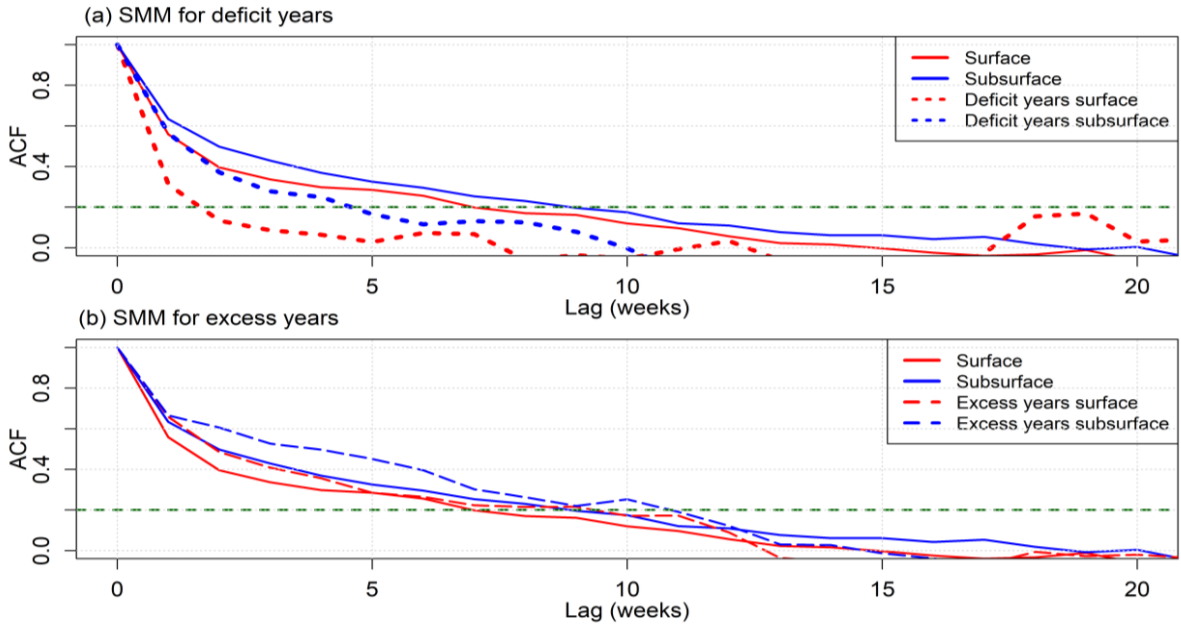

**Figure 5:** Estimates of surface (Red) and subsurface (Blue) SM persistence timescale (Soil moisture memory) during the period of 2000-2021 using 52-week lag auto-correlation function over the CMZ of India. (a) Deficit years are represented by the dotted lines and (b) excess years by the dashed lines. The dashed green line indicates 95% significance level.

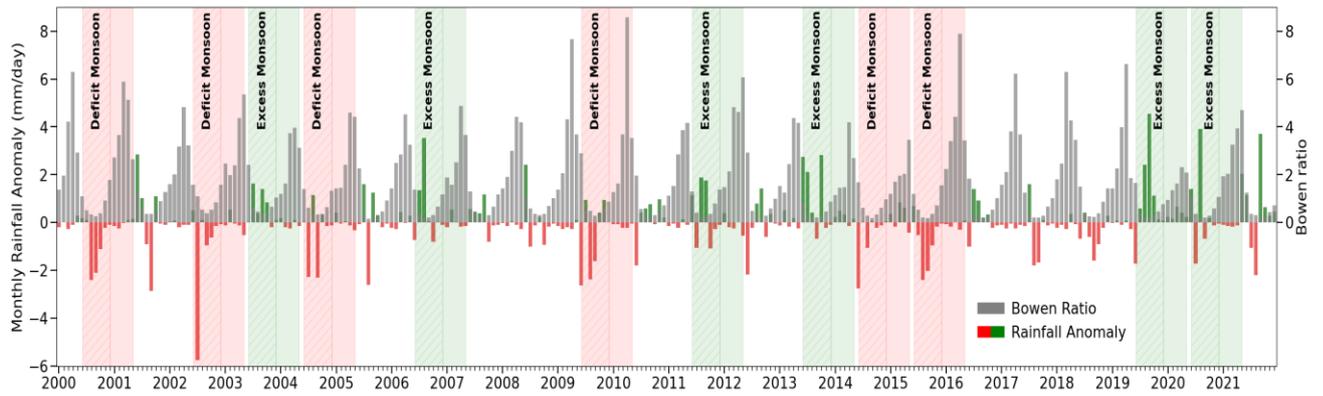

**Figure 6:** The monthly mean rainfall anomaly (mm/day) and Bowen ratio (ratio of sensible to latent heat flux) are shown for the period of 2000-2021 over the CMZ of India. The dark red and green colour bars indicate negative and positive rainfall anomalies respectively, while grey bars indicate Bowen ratio. The hatched region with red (green) shading represents the June-December months of deficit (excess) years. Whereas the plane shaded region with red(green) depicts the successive winter and pre-monsoon season.

**(a) Monthly surface and subsurface soil temperature distribution**

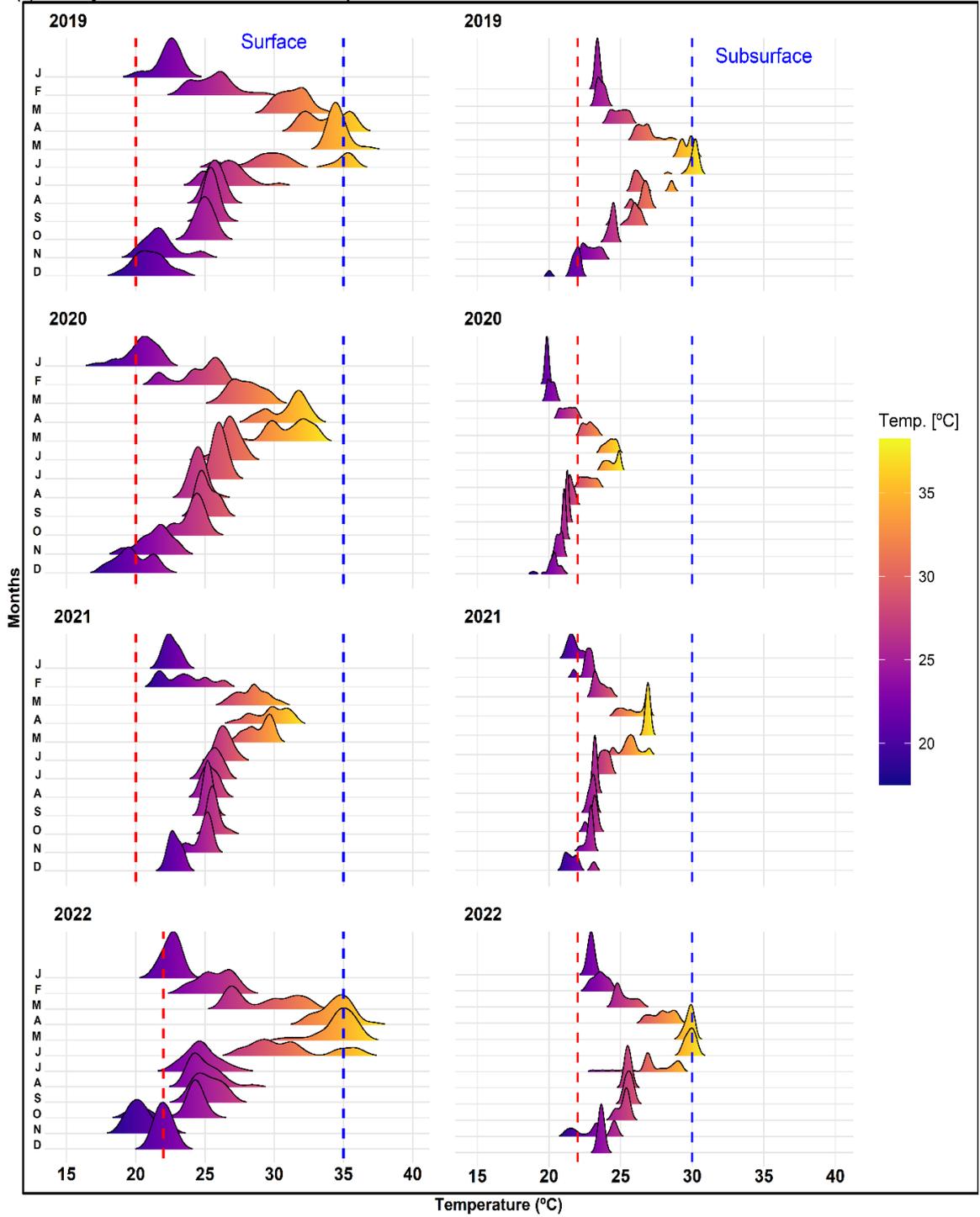

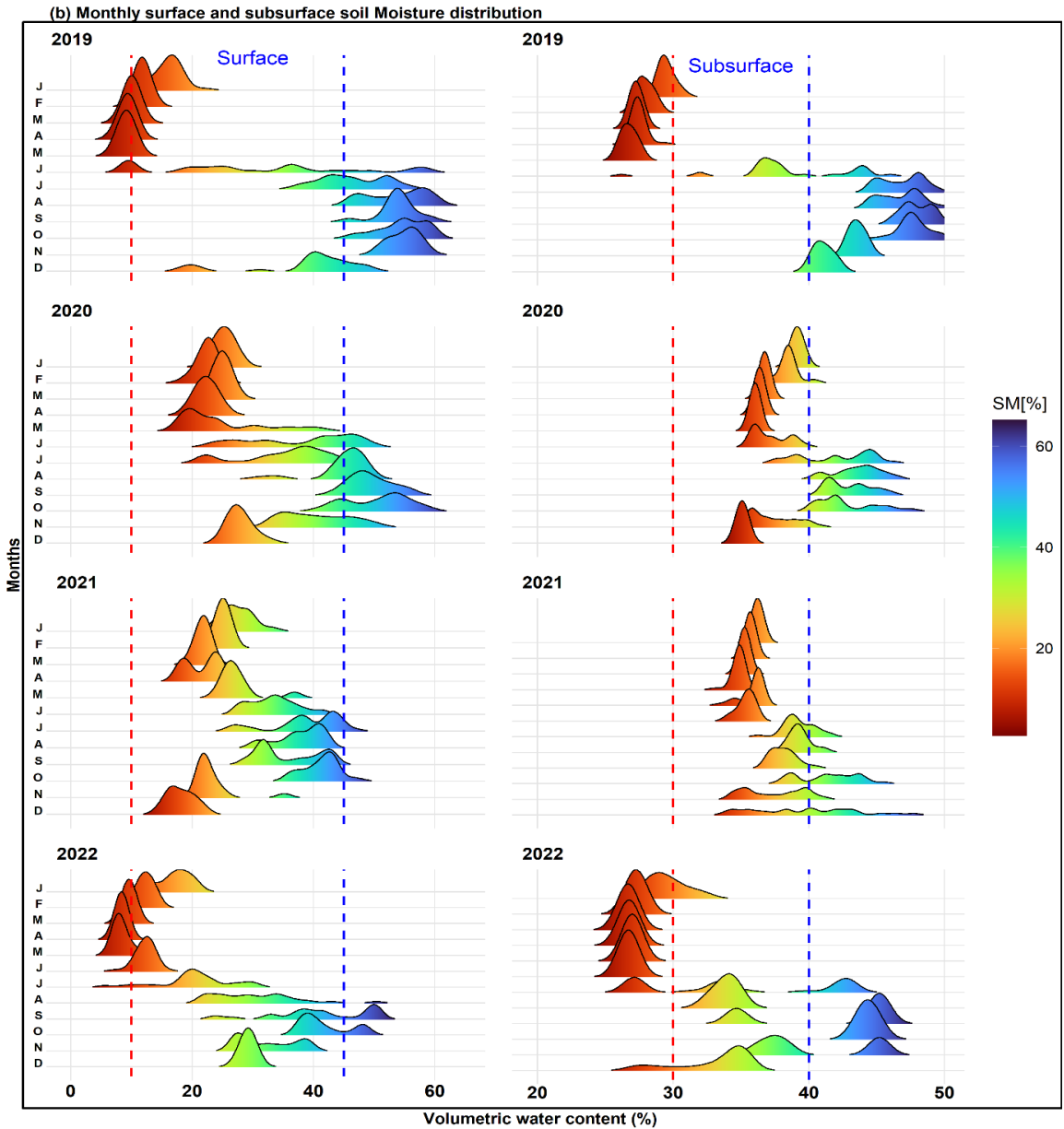

**Figure 7:** The probability density function of surface and subsurface (a) soil temperature (ST, °C) and (b) soil moisture (SM, VWC %) for each month of the period between 2019 and 2022, using in-situ measurements at the COSMOS-IITM, Pune. The dashed red lines indicate the threshold for extremely dry (warm) SM (ST) conditions, whereas the dotted blue line, indicates the threshold for extremely wet (cool) SM (ST) conditions.

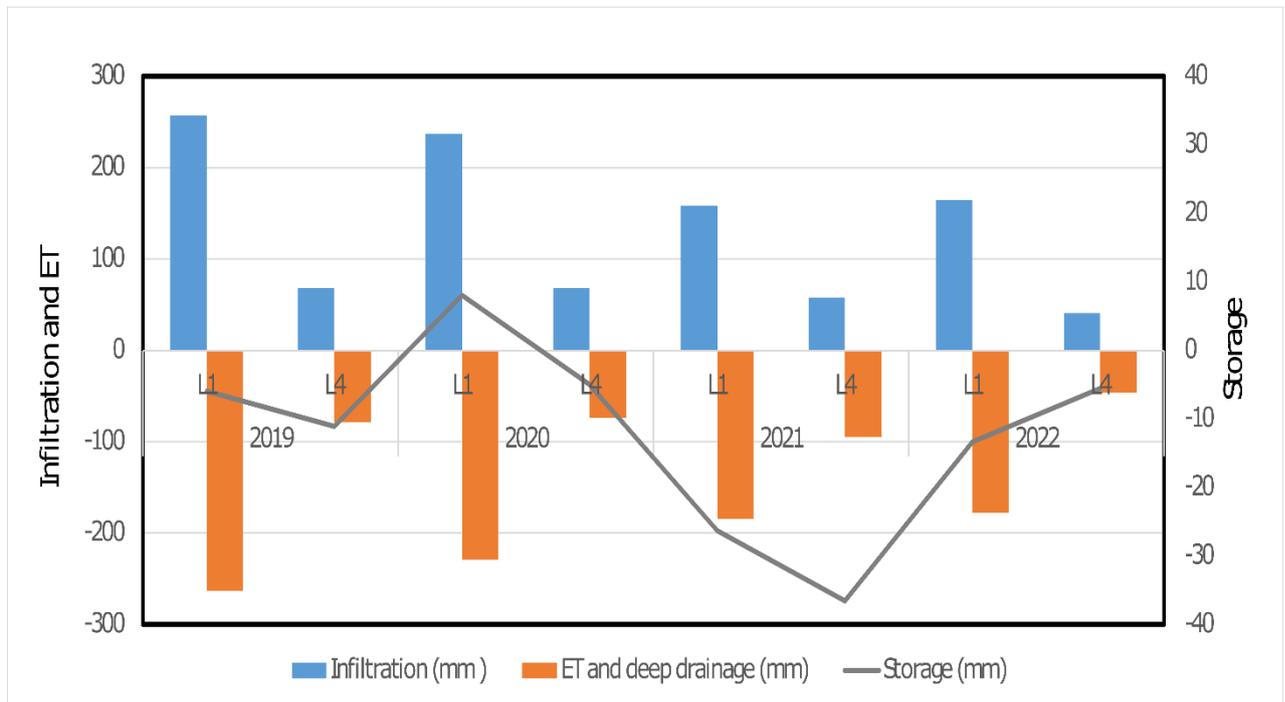

**Figure 8.** The bar plot for net soil water balance at Surface (L1: at the depth of 10 cm) and subsurface (L4: at the depth of 60 cm) levels, for the period of 2019-2022 using daily soil departure and concurrent effective depths. The blue bars represent infiltration, orange ones are for ET and deep drainage and the grey curve depicts storage.

**Supplementary Data:**

**Table S1:** Extreme surface temperature events measured at the COSMOS-IITM site during the period 2017-2021)

| Number of extreme temperature spells (Maximum temperature >= 40 ° C) | 2017 | 2018 | 2019 | 2020 | 2021 |
|---|---|---|---|---|---|
| **Continuously for 1 -2 days** | 3 | 5 | 3 | 0 | 0 |
| **Continuously for 3 days or more** | 4 | 0 | 3 | 0 | 0 |

**Table S2:** Seasonal surface and subsurface SM variability (%) over the COSMOS-IITM site for the period of 2019-2022. Values in red indicate significant seasonal variability during extreme conditions.

| Year | Pre-monsoon | | Monsoon | | Post-monsoon | |
|---|---|---|---|---|---|---|
| | Surface (%) | Subsurface (%) | Surface (%) | Subsurface (%) | Surface (%) | Subsurface (%) |
| 2019 | 8.95 | 13.63 | 58.76 | 20.29 | 19.62 | 8.39 |
| 2020 | 18.08 | 0.85 | 19.79 | 7.59 | 26.02 | 9.63 |
| 2021 | 15.89 | 1.93 | 13.97 | 4.79 | 37.68 | 9.39 |
| 2022 | 10.45 | 18.65 | 46.84 | 19.53 | 18.80 | 9.45 |

**Table S3:** Surface (L1 = depth of 10 cm) and subsurface (L4 = depth of 60 cm) soil water balance estimates, using GLDAS data for the period 2017-2021

| GLDAS Water Budget | 2017 | | 2018 | | 2019 | | 2020 | | 2021 | |
|---|---|---|---|---|---|---|---|---|---|---|
| | L1 | L4 | L1 | L4 | L1 | L4 | L1 | L4 | L1 | L4 |
| Infiltration (mm ) | 130 | 244 | 119 | 144 | 126 | 276 | 134 | 244 | 538 | 240 |
| Evapotranspiration and deep drainage (mm) (all values are in minus) | 126 | 243 | 123 | 151 | 121 | 266 | 134 | 248 | 528 | 240 |
| Storage (mm) (II+III) | 4 | 1 | -4 | -6 | 5 | 10 | 0 | -4 | 10 | 0 |

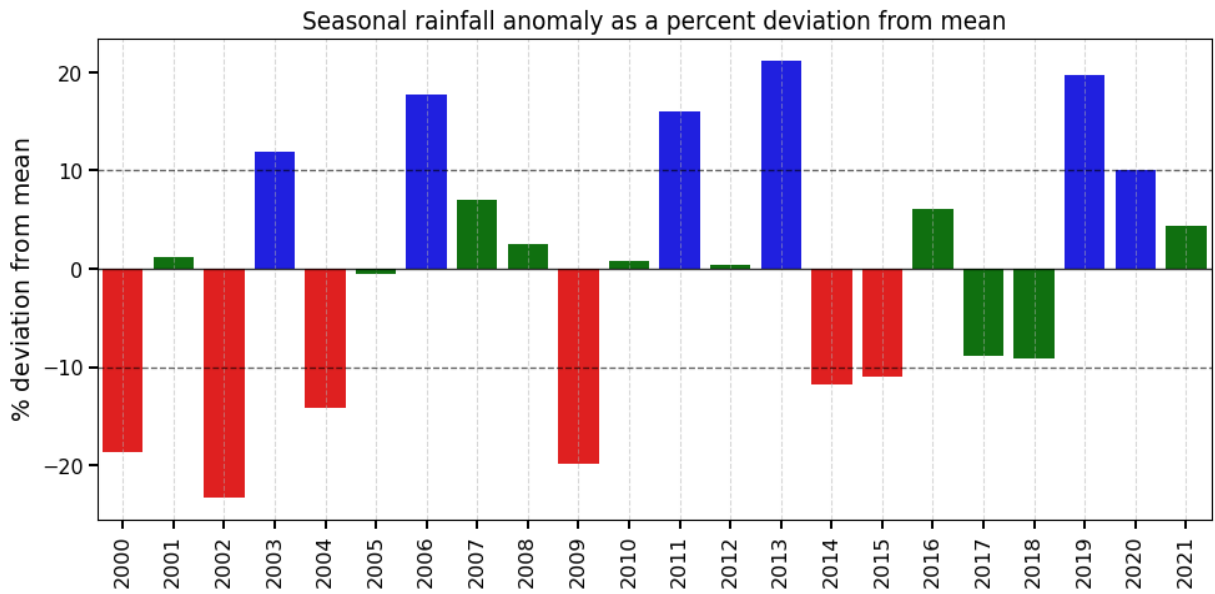

**Figures S1:** Seasonal (June to September) rainfall anomaly expressed as a percent deviation of mean (equation 2, section 2.2.2) over the core monsoon zone of India for the period of 2000-2021. The red and blue bar represents deficit (< -10%) and excess (> 10%) rainfall years, whereas the green bar indicates the normal years.

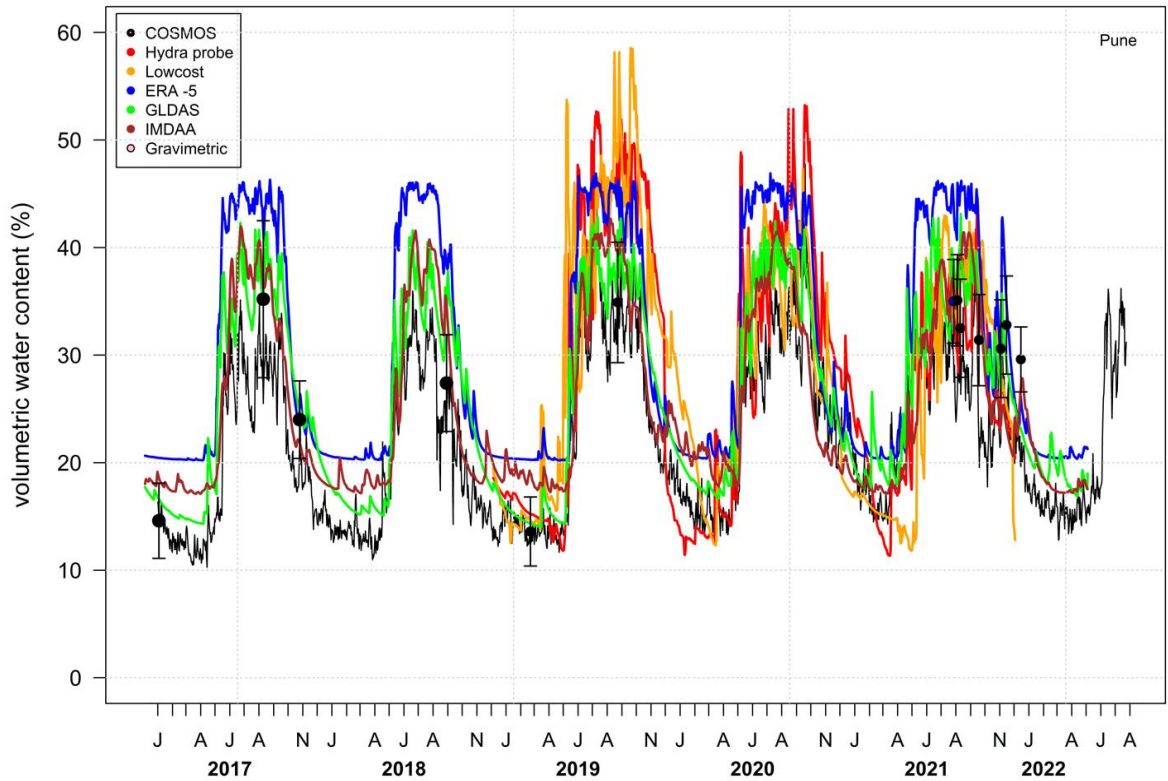

**Figures S2:** Figure shows a daily time series of area-averaged surface SM using various observations (COSMOS, Hydra-probe, low-cost, and Gravimetric), reanalysis (ERA5, IMDAA) and model (GLDAS) data products (represented by distinct colors) for the period of 2017-2022 at COSMOS-IITM, Pune site.

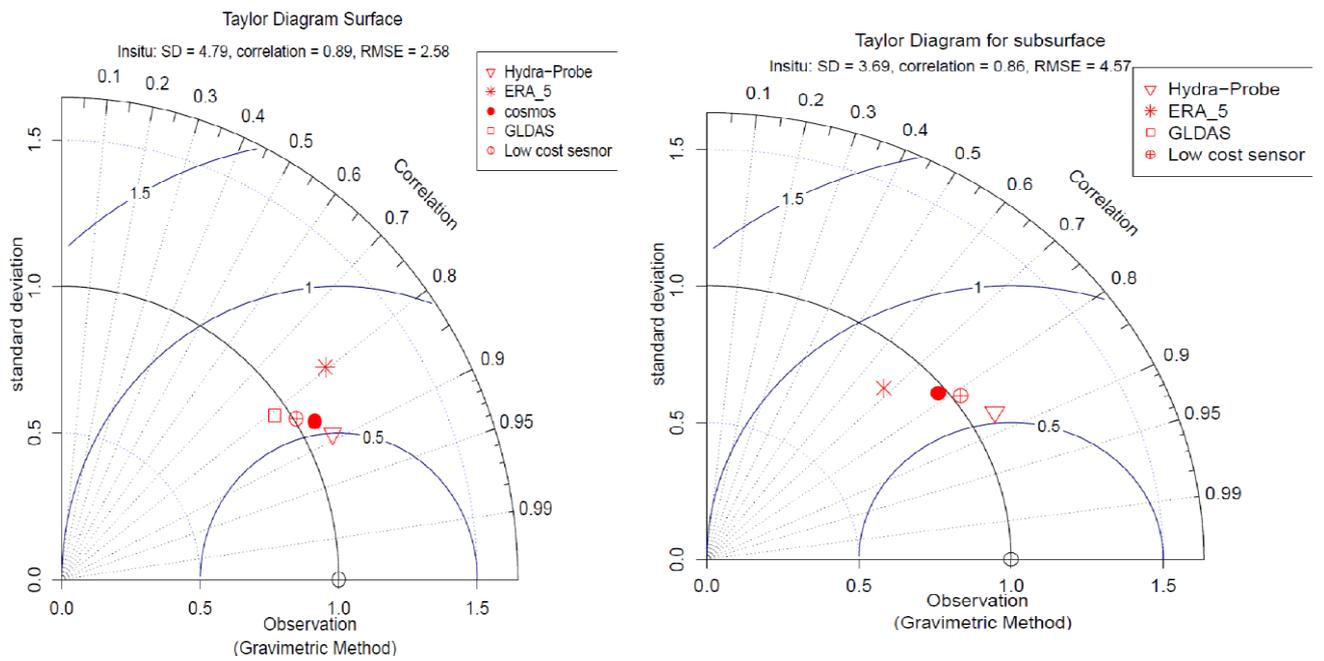

**Figure S3:** Taylor Diagram shows a) surface (10cm) and b) Subsurface (60cm) comparison between different SM data sets (Hydro Probe In-situ sensor, COSMOS, Low-cost sensor, ERA 5, and GLDAS; represented by distinct symbols) with validation reference to gravimetric observation collected at COSMOS-IITM site.

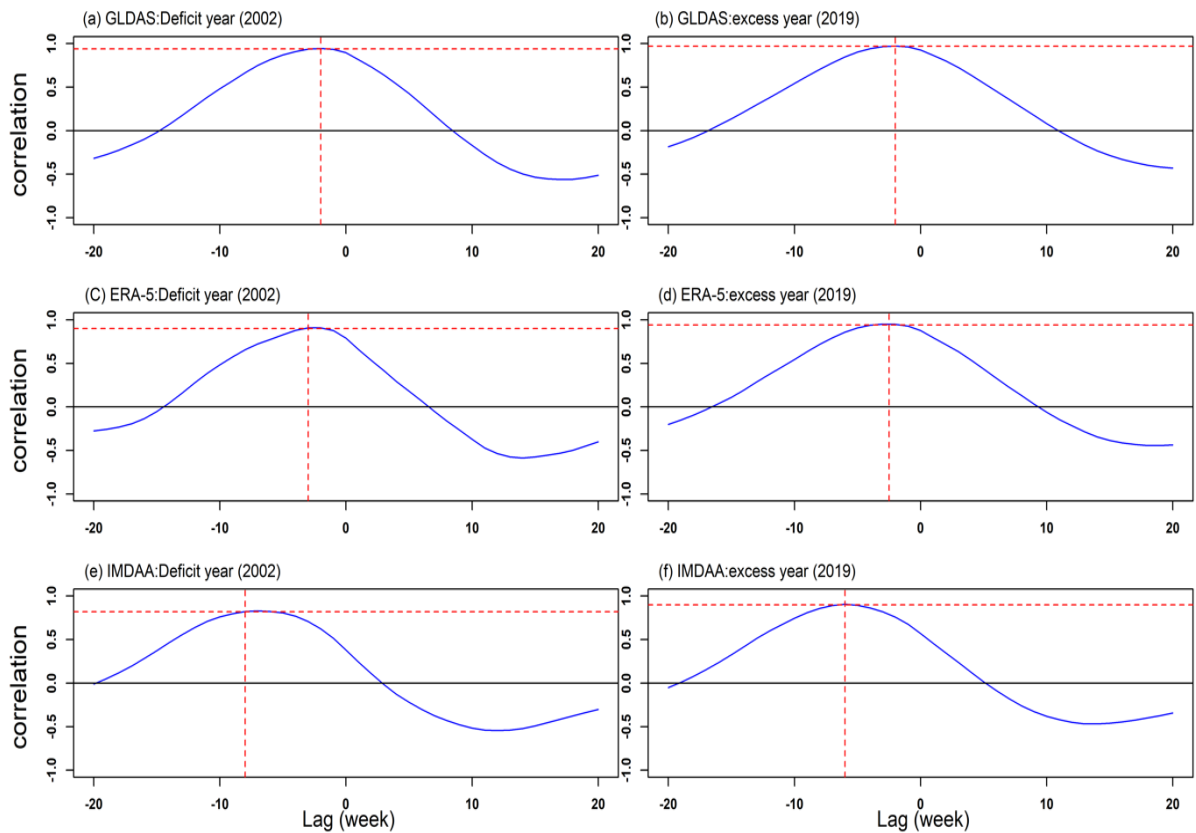

**Figure S4:** Cross correlation between different lags of surface and subsurface SM over the core monsoon zone of India. Panels on the left indicate deficit year (2002) and those on right indicates the excess year (2019). The panels (a) and (b) are for GLDAS, panels (c) and (d) for ERA-5 and panels (e) and (f) for the IMDAA data sets.

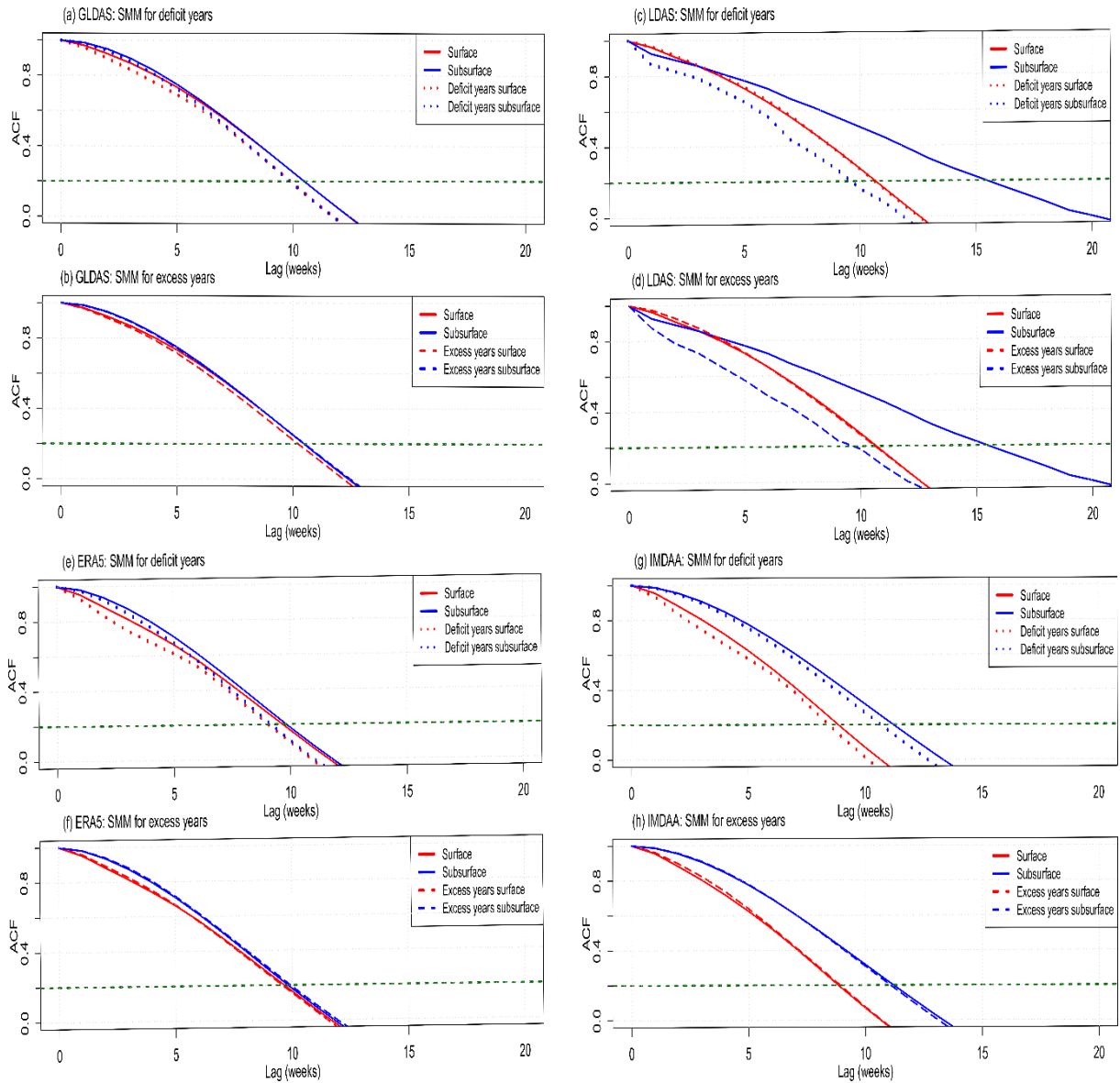

**Figure S5:** Estimates of surface (Red) and subsurface (Blue) SM persistence timescale (Soil moisture memory) during the period of 2000-2021 using 52-week lag auto-correlation function over the CMZ of India. Deficit years (Figure: a, c, e, g) are represented by the dotted lines and excess years (Figure: b, d, f, h) by the dashed lines for GLDAS, LDAS, ERA-5 and IMDAA data sets, respectively. The dashed green line indicates 95% significance level.

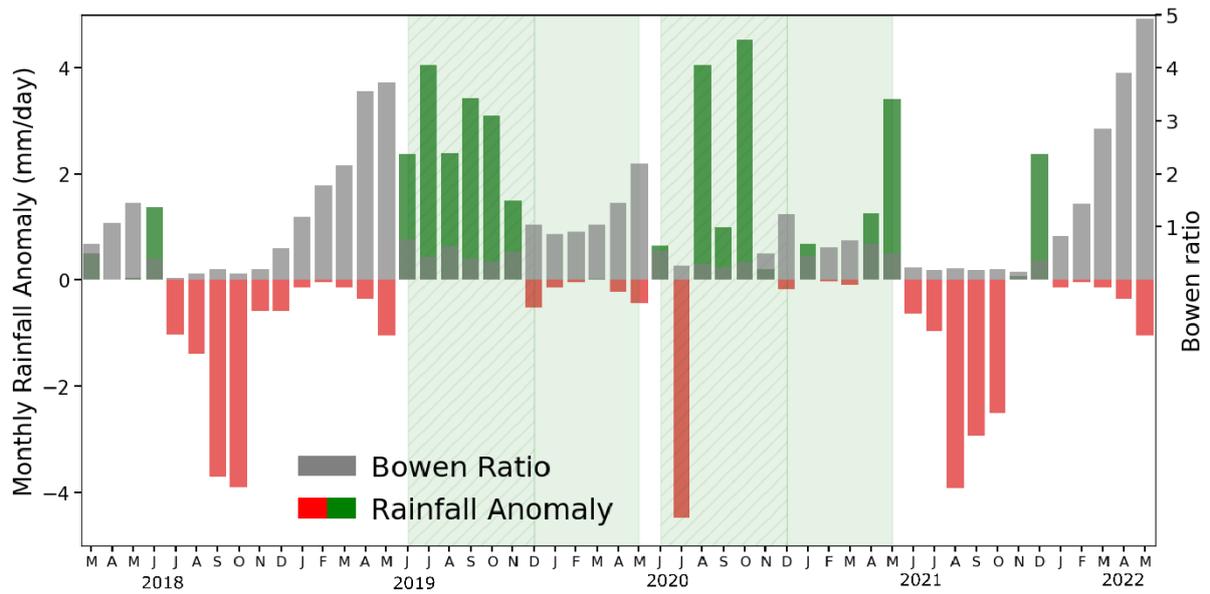

**Figure S6:** Monthly rainfall anomalies (mm/day) indicated as red (negative anomaly) and green (positive anomaly) bar plot. The grey bar plot represents Bowen ratio. The calculations are based on the observations carried out the COSMOS-IITM, Pune site during 2019̶8 to 2021. The hatched green region represents the June-December months of an excess year, whereas the plane shaded region shows the successive winter and pre-monsoon season.

**Declaration of interests**

The authors declare that they have no known competing financial interests or personal relationships that could have appeared to influence the work reported in this paper.

**Acknowledgments**

The authors are grateful to the Director, Indian Institute of Tropical Meteorology (IITM, India), for his unconditional support to carry out this research work. IITM is an autonomous research Institute, fully funded by the Ministry of Earth Sciences (MoES), Govt. of India. The IITM HPC support is duly acknowledged. The available Hydro-meteorological data sets from Flux Tower measurements, and India Meteorological Department (IMD) observations have been used for cross-validation and are acknowledged here. The logistic support provided by the IITM for maintaining this COSMOS-IITM observational site is duly acknowledged.

**Author Contribution:**

MMG initiated and led this study. MM, RK, BBS conceived and supervised the study. MMG performed the analyses supported by MI, BBS, and NG. MMG, MM, BBS, MI wrote the original draft, reviewed by RK, SNP, TF, DN, PS, and MR. All authors participated in discussions and contributed to review and editing the manuscript.